\def\aj{{AJ}}

\def\apj{{ApJ}}

\def\mnras{{MNRAS}}

\def\asec{$^{\prime\prime}$}

\documentclass[cup5b]{caps}
\usepackage{psfig}
\usepackage{graphicx}
\usepackage{amssymb}
\usepackage{ociwsymp1}
\HeadText{J. S. Dunlop}

\def\plotone#1{\centering \leavevmode
\includegraphics[width=.95\columnwidth]{#1}}

\begin{document}

\pagenumbering{arabic}

\author[]{JAMES S. DUNLOP\\Institute for Astronomy, 
University of Edinburgh, Royal Observatory}

\chapter{Quasar Hosts and the Black Hole-Spheroid Connection}

\begin{abstract}
I review our current understanding of the structures and ages of the host 
galaxies of quasars, and the masses of their central black holes. At low 
redshift, due largely to the impact of the {\it Hubble Space Telescope}, there 
is now compelling evidence that the hosts of quasars with $M_V < -24$ mag are 
virtually all massive ellipticals, with basic properties indistinguishable 
from those displayed by their quiescent counterparts. The masses of these 
spheroids are as expected given the relationship between black hole and 
spheroid mass now established for nearby galaxies, as is the growing 
prevalence of significant disk components in the hosts of progressively 
fainter active nuclei. In fact, from spectroscopic measurements of the 
velocity of the broad-line region in quasars, it has now proved possible to 
obtain an independent dynamical estimate of the masses of the black holes 
that power quasars. I summarize recent results from this work, which can be 
used to demonstrate that the black hole-spheroid mass ratio in quasars is the 
same as that found for quiescent galaxies, namely $M_\bullet = 0.0012 
M_{\rm sph}$.
These results offer the exciting prospect of using observations of quasars 
and their hosts to extend the study of the black hole-spheroid mass ratio out 
to very high redshifts ($z > 2$). Moreover, there is now good evidence that 
certain ultraviolet quasar emission lines can provide robust estimates of 
black hole masses from the observed optical spectra of quasars out to 
$z > 2$, and perhaps even at $z > 4$. By combining such information with 
deep, high-resolution infrared imaging of high-redshift quasar hosts on 8-m 
class telescopes, there is now a real prospect of clarifying the evolution 
of the black hole spheroid connection over cosmological time scales.
\end{abstract}

\section{Introduction}

With the discovery that all spheroids (i.e., elliptical 
galaxies and disk galaxy bulges) appear to house a 
massive black hole of proportionate mass (Magorrian et al. 1998; 
Gebhardt et al. 2000a; Merritt \& Ferrarese 2001),
the nature and evolution of quasar host 
galaxies has grown from a subject explored primarily by
AGN researchers, into an area of interest for all astronomers
concerned with the formation and evolution of galaxies and 
of compact objects.

Indeed, black hole and spheroid formation/growth are now 
recognized as potentially intimately related processes (Silk \& Rees
1998; Fabian 1999; Granato et al. 2001; Archibald et al. 2002), 
with the evolution of quasar host galaxies as a function of redshift 
now seen as a key measurement in observational cosmology (e.g.,
Kauffmann \& Haehnelt 2000). In this review I have chosen to focus on what 
can be learned about the nature of the black hole spheroid connection from 
observations of quasars and their hosts. I have therefore deliberately 
avoided detailed discussion of many other topics of interest related 
to quasar host galaxy research, such as the triggering of quasar activity,
the origin of radio loudness, and the nature of possible links between
quasars and ultraluminous infrared galaxies (ULIRGs). 

This Chapter is divided into three sections as follows. 
First I summarize what has been 
learned about the host galaxies of low-redshift quasars from deep 
imaging/spectroscopy over the last decade. Second I discuss the latest 
dynamical estimates of the masses of the black holes that power quasars.
Third I consider the immediate future prospects for extending these two 
prongs of measurement to higher redshift, to explore the nature of the 
black hole spheroid connection as a function of cosmological time.

Unless otherwise stated, an Einstein-de Sitter Universe with $H_0 = 50\,{\rm 
km s^{-1} Mpc^{-1}}$ has been assumed for the calculation of physical 
quantities.
 
\section{The Host Galaxies of Low-redshift Quasars}

Many imaging studies and several spectroscopic studies 
of ``nearby'' ($z < 0.3$) quasar host galaxies have been attempted 
over the last quarter of a century, but it is only in the last decade 
that a clear picture of the nature of quasar hosts has emerged from this work.
This progress can be attributed first to the advent of deep near-infrared 
imaging (Dunlop et al. 1993; McLeod \& Rieke 1994; Taylor et al. 1996),
and second to the high angular resolution provided by the refurbished
{\it Hubble Space Telescope (HST)}\ (e.g., Disney et al. 1995; Bahcall et al. 
1997; Hooper, Impey, \& Foltz 1997; McLure et al. 1999). 

Some workers have chosen to focus on some of the morphological peculiarities
and evidence of ``action'' revealed by this deep imaging, such as tidal tails, 
and nearby companions (perhaps responsible for triggering 
the nuclear activity). However, the clearest results, and most meaningful 
insights have emerged from studies that have focused on determining the 
properties of the mass-dominant stellar populations in quasar hosts, and 
exploring how these compare with those of quiescent galaxies.

Figure 1.1 provides an example of how clearly the basic structure of 
low-redshift quasar host galaxies can be discerned with an exposure of 
$\sim 1$ hour on the {\it HST}. This image (taken from Dunlop et al. 2003) 
demonstrates not only that the host galaxy is well resolved, but also 
the extent to which the vast majority of the optical light from the 
host can generally be attributed to a simple, symmetric, ``normal'' galaxy
(in this case an elliptical, with an 
$r^{1/4}$ de Vaucouleurs luminosity profile, and a half-light radius
of 7.5 kpc).

For simplicity I have chosen to center the following summary of what has 
been learned from such images around the main results from our 
own, recently completed,
{\it HST}\ imaging study of the hosts of radio-loud quasars (RLQs), radio-quiet
quasars (RQQs) and radio galaxies (RGs) at $z \simeq 0.2$
(Dunlop et al. 2003). However, wherever appropriate, I have also 
endeavored to discuss (and if possible explain) the extent to which other 
authors do or do not agree with our findings.

\begin{figure}
\plotone{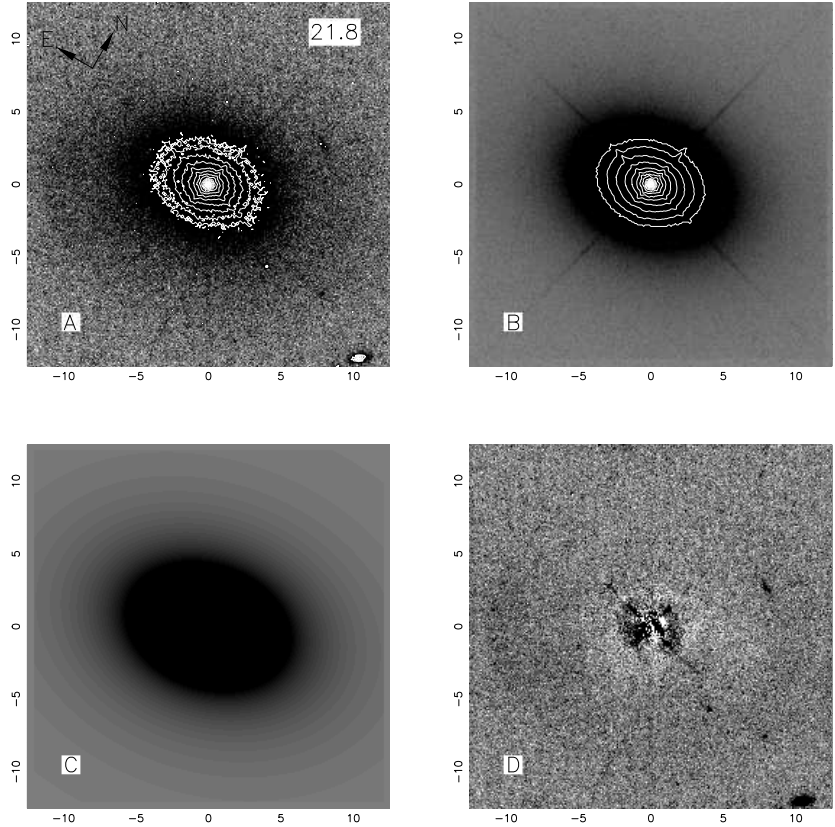}
\caption{An example of deep {\it HST}\ imaging of the host galaxy of a 
low-redshift quasar. A greyscale/contour representation of an $R$-band 
image of the $z = 0.1$ RQQ 0204$+$292 obtained with the WFPC2 is shown in the 
upper-left panel (the image is 25\asec$\times$ 25\asec\ in size). The 
upper-right panel shows the best-fitting model of this image (after 
convolution with the {\it HST}\ point-spread function), which comprises a 
de Vaucouleurs law  elliptical galaxy (of half-light radius $r_{1/2} = 
7.5\,{\rm kpc}$) along with an unresolved nuclear component.  The lower-left 
image shows the best-fitting host galaxy as it would appear if the nucleus 
were inactive, while the lower-right panel is the residual image
that results from subtraction of the complete model from the image.
Further details of the modeling procedure used can be found in McLure, Dunlop, 
\& Kukula (2000) and Dunlop et al. (2003).}
\end{figure}

\subsection{Host Galaxy Luminosity, Morphology, and Size}

\begin{figure}
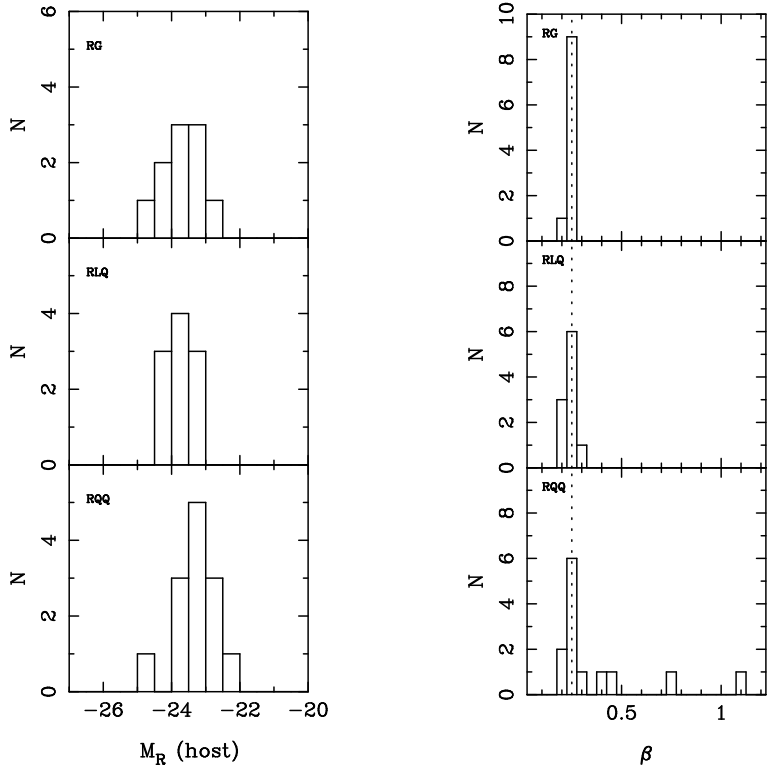

\psfig{file=fig2anew.ps,width=4.25cm,angle=0}     
\vspace*{-10.20cm}
\hspace*{6cm}
\psfig{file=fig2bnew.ps,width=4.0cm,angle=0}    
\caption{{\it Left:}\ Histograms of host galaxy integrated $R$-band absolute 
magnitudes for the RG, RLQ, and RQQ subsamples imaged with the {\it HST}\ by 
Dunlop et al. (2003). For comparison, the integrated $R$-band absolute
magnitude of an $L^{\star}$ galaxy is $M_R = -22.2$ mag. {\it Right:}\ 
Histograms of the best-fit values of $\beta$, where 
host galaxy surface brightness is proportional to exp$(-r^{\beta})$, for 
the same three subsamples. The dotted line at $\beta=0.25$
indicates a perfect de Vaucouleurs law, and all of the radio-loud hosts
and all but three of the radio-quiet hosts are consistent with this to 
within the errors. Two of the three RQQs with
hosts for which $\beta > 0.4$ transpire to be the two least luminous 
nuclei in the sample, and should be reclassified as Seyferts.}
\end{figure}

After some initial confusion (e.g., Bahcall, Kirhakos, \& Schneider 1994), recent {\it HST}-based
studies have now reached agreement that the hosts of all luminous quasars 
($M_V < -23.5$ mag) are bright galaxies with $L > L^{\star}$ (McLure et al. 1999;
McLeod \& McLeod 2001; Dunlop et al. 2003). This result is illustrated 
by Figure 1.2 (left panel), taken from Dunlop et al. (2003).
However, it can be argued, with justification, that this much had 
already been established from earlier ground-based studies 
(e.g., Taylor et al. 1996). 

In fact the major 
advance offered by the {\it HST}\ for the study of quasar hosts is that it has 
enabled host luminosity profiles to be measured over sufficient angular and 
dynamic range to allow a de Vaucouleurs $r^{1/4}$-law spheroidal component to 
be clearly distinguished from an exponential disk, at least for redshifts 
$z < 0.5$. 

In our own study 
this is the reason that we have been able to establish unambiguously 
that, at low $z$,  the hosts of both RLQs {\it and}\
RQQs are undoubtedly 
massive ellipticals with (except for one RQQ in our sample) 
negligible disk components 
(McLure et al. 1999; Dunlop et al. 2003).
This result is illustrated in the right panel of Figure 1.2. 
This figure confirms that the hosts of RQQs and RGs all
follow essentially perfect de Vaucouleurs profiles, in good agreement with 
the results of other studies. Perhaps the more surprising aspect 
is the extent to which the RQQ sample is also dominated 
by spheroidal hosts. At first sight this might seem at odds with the results
of some other recent studies, such as those of Bahcall et al. (1997) and 
Hamilton, Casertano, \& Turnshek (2002), who report that 
approximately one-third to one-half of RQQ lie in 
disk-dominated hosts.
However, on closer examination, it becomes clear that there is no real 
contradiction provided one compares quasars of similar power. 
Specifically, if attention is confined to quasars with nuclear magnitudes 
$M_V < -23.5$ mag we find that 10 out of the 11 RQQs in our sample lie in 
ellipticals,
6 out of 7 similarly luminous quasars in the sample of Bahcall et al. lie in 
ellipticals, and 
at least 17 out of the 20 comparably luminous RQQs in Hamilton et al.'s 
archival sample also appear to lie in spheroidal hosts. 

It is thus now clear that above a given luminosity threshold we enter a regime
in which AGNs can only be hosted by massive spheroids, regardless 
of radio power (a result confirmed by the recent {\it HST}\ study of 
the most luminous low-redshift quasars by Floyd et al. 2003). 
It is also clear that, within the radio-quiet
population, significant disk components become more common at lower
nuclear luminosities. This dependence of host galaxy morphology on nuclear 
luminosity is nicely demonstrated by combining our own results with those of 
Schade, Boyle, \& Letawsky (2000) who have studied the host galaxies of lower-luminosity
X-ray selected AGNs. This is shown in Figure 1.3 where the ratio
of bulge to total host luminosity is plotted as a function of nuclear 
optical 
power. Figure 1.3 is at least qualitatively as expected if black hole mass is 
proportional to spheroid mass, and black hole masses $> 2 \times 10^8\,
M_{\odot}$ are required to produce quasars with $M_R < -23.5$ mag.

\begin{figure}
\plotone{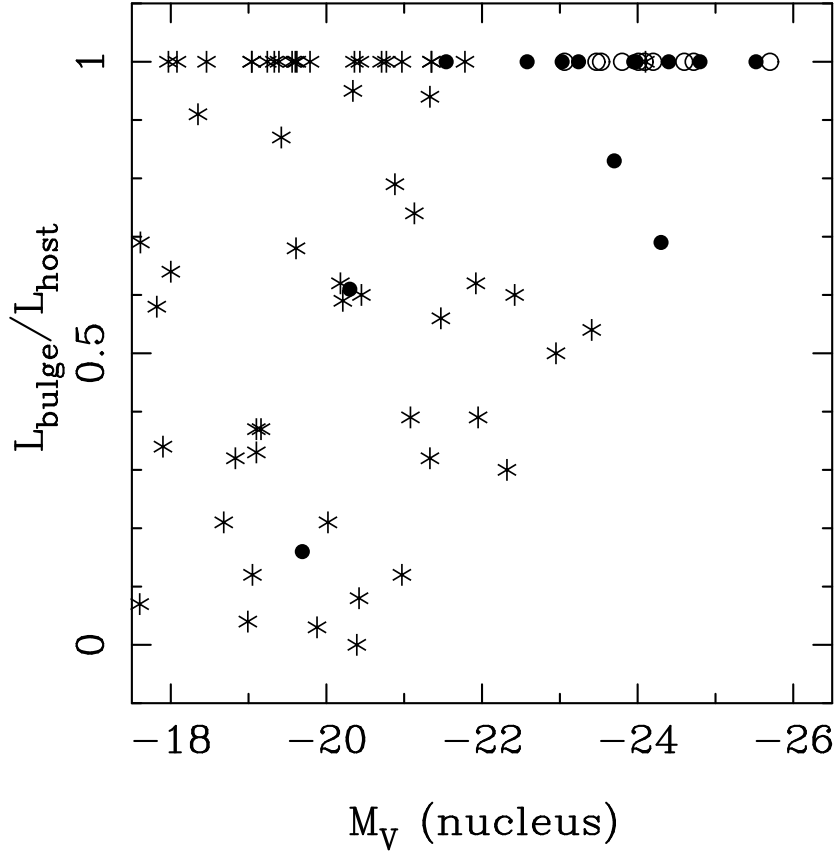}
\caption{The relative contribution of the spheroidal
component to the total luminosity of the host galaxy plotted against the 
absolute
$V$-band magnitude of the nuclear component. The plot shows the results
from Dunlop et al. (2003) (RLQs as open circles, RQQs as filled circles)
along with the results from Schade et al. (2000) for a larger sample of
X-ray selected AGNs spanning a wider but lower range of 
optical luminosities (asterisks). 
This plot illustrates very clearly how disk-dominated
host galaxies become increasingly rare with increasing nuclear power, as
is expected if more luminous AGNs are powered by more massive black holes,
which, in turn, are housed in more massive spheroids.}
\end{figure}

In our {\it HST}\ study we have also been able to break the well-known degeneracy
between host galaxy surface brightness and size. This point is illustrated
by the fact that we have, for the first time, been able to demonstrate
that the hosts of RLQs and RQQs follow Kormendy's (1977) relation (i.e., 
the photometric projection of the fundamental plane; Fig. 1.4). 
Moreover the  
slope ($2.90 \pm 0.2$) and normalization of this relation are 
identical to that displayed by normal quiescent massive ellipticals.

\begin{figure}
\includegraphics[width=.95\columnwidth,angle=270]{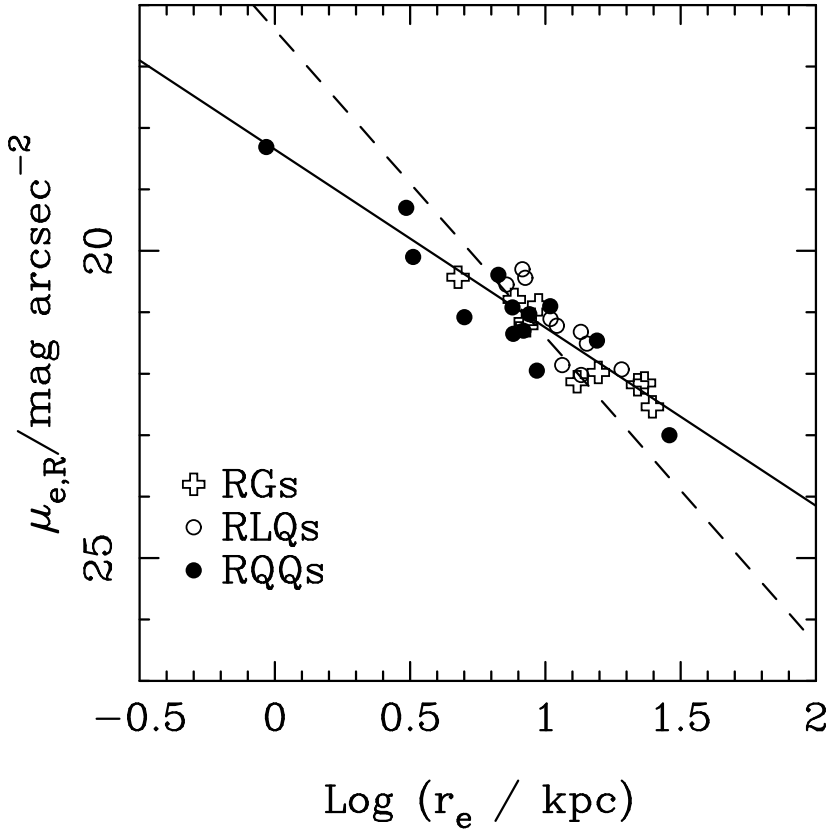}
\caption{The Kormendy (1977) relation followed by the hosts of all 33 powerful 
AGNs studied by Dunlop et al. (2003) with the {\it HST}. The solid line is the 
least-squares fit to the data that has a slope of $2.90 \pm 0.2$, in 
excellent agreement with the slope of 2.95 found by Kormendy for 
inactive ellipticals. For the few RQQs that have a disk component the 
best-fitting bulge component has been plotted. The dashed line has a slope of
5, indicative of the ``pseudo Kormendy relation'' expected if  
the scale lengths of the host galaxies had not been properly constrained 
(see Dunlop et al. 2003).}
\end{figure}

\subsection{Host Galaxy Ages}

It is well known from simulations that the merger of two disk galaxies can 
produce  a remnant that displays a luminosity profile not dissimilar
to a de Vaucouleurs $r^{1/4}$ law. This raises the possibility that the 
apparently spheroidal nature of the quasar hosts discussed above might be
the result of a recent major merger that could also be responsible for 
stimulating the onset of nuclear activity. This would also be the natural
prediction of suggested evolutionary schemes in which ULIRGs are 
presumed to be the precursors of RQQs. Could a recent merger of two massive,
gas-rich disks be simultaneously responsible for the triggering of 
nuclear activity and the production of an apparently spheroidal host?

The answer, at least at low redshift, appears to be no. 
First, as mentioned above, the Kormendy relation
displayed by quasar hosts appears to be indistinguishable from that
of quiescent, well-evolved massive ellipticals. Moreover, as discussed
by Genzel et al. (2001) and Dunlop et al. (2003), ULIRGs generally lie 
in a different region of the fundamental plane to quasar hosts, with the former
apparently destined to evolve into lower or intermediate-mass spheroidal
galaxies.

Secondly, direct attempts to determine the ages of the dominant stellar
populations in the quasar hosts provide little evidence of recent,
widespread star formation activity. Within the Dunlop et al.
sample we have attempted to estimate the ages of the host galaxies 
both from optical-infrared colors and from deep optical off-nuclear 
spectroscopy (Nolan et al. 2001). The 
results of this investigation are that the hosts of both radio-loud and 
radio-quiet quasars are dominated by old, well-evolved stellar populations 
(with typically less than 1\% of stellar mass involved in recent 
star formation activity).
There are currently no comparably extensive studies of 
host galaxy stellar populations with which 
this result can be compared. However,
Canalizo \& Stockton (2000) have published results from a more detailed 
spectroscopic study of three objects, one of which, Mrk 1014,
is also in the Dunlop et al. RQQ sample. This is in fact the only quasar host
in the Dunlop et al. sample for which Nolan et al. (2001)
found clear spectroscopic evidence of A-star 
features and a significant (albeit still only $\simeq 2$\% by mass) 
young stellar population. It 
is presumably no coincidence that this is also the only quasar in the 
Dunlop et al. sample that was detected by {\it IRAS}, and the only host that
displays spectacular tidal tail features comparable to those commonly 
found in images of ULIRGs. However, even for this apparently
star-forming quasar host, Canalizo \& Stockton agree that $\simeq 95$\% of 
the host is dominated by an old, well-evolved stellar population (although 
they argue that 5\%--8\% of the galaxy has been involved in recent star 
formation). 

Finally, despite claims to the contrary, recent measurements of molecular gas
in AGN host galaxies reported by Scoville et al. (2003) are completely
consistent with this picture. Scoville et al. detected substantial molecular
gas masses in the hosts of lower luminosity quasars with known 
substantial disk components,
and failed to detect molecular gas in the hosts of the
three most luminous quasars in their sample.

In summary, the available evidence indicates that the hosts of quasars 
with $M_V < -23.5$ mag are virtually all massive elliptical galaxies. Moreover, 
quasar hosts appear to be ``normal'' ellipticals in the sense that their basic 
structural properties, and the ages of their dominant stellar populations 
are, at least to first order, indistinguishable from those of their 
quiescent counterparts. Both the universality of elliptical hosts for the 
most luminous low-redshift quasars and the growing prevalence of significant 
disk components in the hosts of progressively fainter active nuclei can be 
viewed as a natural reflection of the proportionality of black hole and 
spheroid mass now established for nearby quiescent galaxies. In the next 
section I describe the results of recent attempts to obtain dynamical 
estimates of the masses of the black holes that power quasars. Such studies 
allow a direct test of whether or not the constant of proportionality
between black hole and spheroid mass is the same in the active and inactive 
galaxy populations.

\section{The Black Hole-Spheroid Mass Ratio in Low-redshift Quasars}

If one assumes that quasars emit at the Eddington limit, it is straightforward
to obtain a very rough estimate of the masses of their central black holes. 
However, a potentially much more reliable estimate 
of black hole mass can be obtained via an analysis of the velocity
widths of the H$\beta$ lines in quasar nuclear spectra.
This has been a growth industry in recent years (e.g., Wandel 1999; 
Laor 2000), bolstered by estimates of the size of the broad-line region (BLR)
from reverberation mapping of low-redshift, broad-line AGNs. 

\subsection{The Virial Black Hole Mass Estimator}

The underlying assumption behind the virial black hole mass estimator is that 
the motion of the broad-line emitting material in AGNs is virialized. Under 
this assumption the width of the broad lines can be used to trace the 
Keplerian velocity of the broad-line gas, and thereby allow an estimate of the 
central black hole mass via the formula 
$M_\bullet=G^{-1}R_{\rm BLR}V_{\rm BLR}^{2}$, where $R_{\rm BLR}$ is the 
BLR radius and $V_{\rm BLR}$ is the
Keplerian velocity of the BLR gas. Currently, the most direct measurements 
of the central black hole masses of powerful AGNs are for 17 Seyferts and 
17 PG quasars for which reverberation mapping has provided a direct
measurement of $R_{\rm BLR}$ (Wandel, Peterson \& Malkan 1999; Kaspi et 
al. 2000). 

An important outcome from these studies is the discovery of a correlation 
between $R_{\rm BLR}$ and the monochromatic AGN continuum luminosity at 5100 
\AA\, (e.g., $R_{\rm BLR}\propto \lambda L_{5100}^{0.7}$; Kaspi et al. 2000). 
By combining this luminosity-based $R_{\rm BLR}$ estimate with a measure
of the BLR velocity based on the FWHM of the H$\beta$ emission line, 
it is now possible to produce a virial black hole mass estimate from a 
single spectrum covering H$\beta$. This technique has recently been widely 
employed to investigate how the masses of quasar black holes relate to the
properties of the surrounding host galaxies (e.g., Laor 2001;  McLure \& 
Dunlop 2001,2002) and the radio luminosity of the central engine (Lacy et al. 
2001; Dunlop et al. 2003).

Of most direct relevance to the topic of interest in this review is the 
result shown in Figure 1.5. This shows how the relationship between host galaxy 
luminosity and black hole mass derived for quasars and Seyferts compares with 
that derived for normal galaxies (McLure \& Dunlop 2002).

Under the assumption that $M_\bullet \propto M_{\rm sph}$ the
best-fitting constant of proportionality derived from the fit to the
quasar and Seyfert data points in Figure 1.5 is 0.0012. This is essentially
identical to the value (0.0013) for nearby inactive galaxies derived by
Kormendy \& Gebhardt (2001), and to the value (0.0012) derived by
Merritt \& Ferrarese (2001). While the virtually exact agreement between
these numbers may be fortuitous, the similarity of the mass relationships 
derived for the active and inactive samples can be fairly viewed as providing 
confirmation both that the $M_\bullet - M_{\rm sph}$ relation is the same in 
active and inactive galaxies, and that the assumption of gravitational 
equilibrium made in applying the H$\beta$ virial mass estimator is valid (see 
also Gebhardt et al. 2000b).

\begin{figure}
\psfig{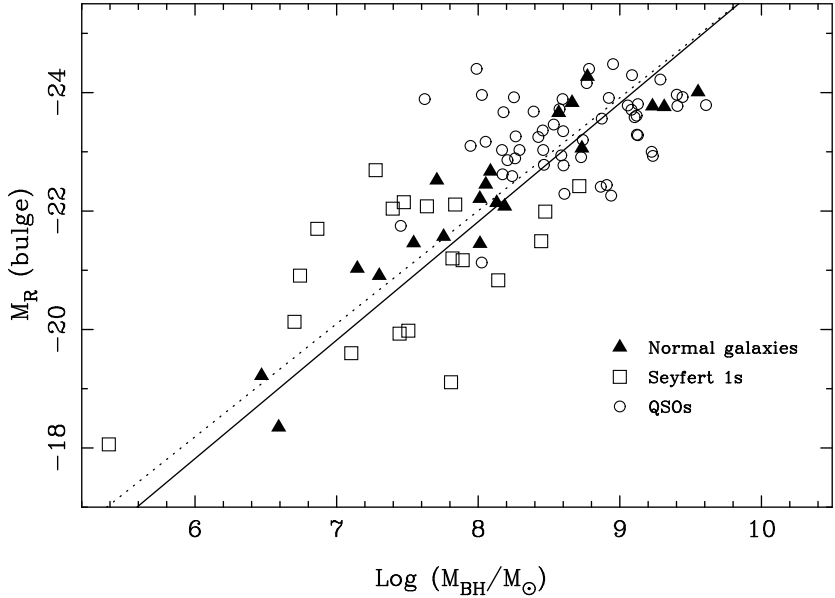}     
\caption{Absolute $R$-band bulge magnitude versus black hole mass plotted for 
72 AGNs and 18 inactive elliptical galaxies. The black hole masses for the 72 
AGNs are derived from their H$\beta$ line widths using a disklike BLR model 
(see McLure \& Dunlop 2002). The black hole masses of the inactive galaxies 
(triangles) are dynamical estimates as compiled by Kormendy \& Gebhardt 
(2001). Also shown is the formal best fit (solid line) and the best-fitting
linear relation between spheroid and black hole mass (dotted line).}
\end{figure}

\subsection{Eddington Ratios}

Having confirmed the constant of proportionality between host spheroid and 
black hole mass for quasars one can then re-address the issue of how the 
actual nuclear luminosities of quasars compare with their predicted 
Eddington-limited values (as inferred from the luminosities of their host 
galaxies).

This is illustrated in Figure 1.6, in which host galaxy absolute $V$-band 
magnitude is plotted against quasar nuclear absolute magnitude for an expanded
sample of quasars assembled from five recent studies [see figure caption
and Floyd et al. (2003) for details]. Also shown in this plot are the 
predicted relations for black holes emitting at 100\%, 10\% and 1\% of the 
Eddington limit.

This plot provides (perhaps surprisingly good) evidence that, at any given 
host luminosity, the most luminous quasar nuclei are emitting at the 
predicted Eddington limit (as calculated on the basis of a black hole mass
inferred from host luminosity using the relation shown in Fig. 1.5). It also 
shows that the majority of low-redshift quasars studied to date are emitting 
at between 10\% and 100\% of the Eddington limit, and that their host galaxies 
range in luminosity from $L^{\star}$ to 10 $L^{\star}$.

In concluding this section, I note that on the basis of this plot it would be 
predicted that the most luminous quasars found in the high-redshift
Universe, with $M_V < -27$ mag, 
can only be produced by the black holes at the centers of the most massive 
(10$L^{\star}$) ellipticals, or their progenitors.

\begin{figure}
\plotone{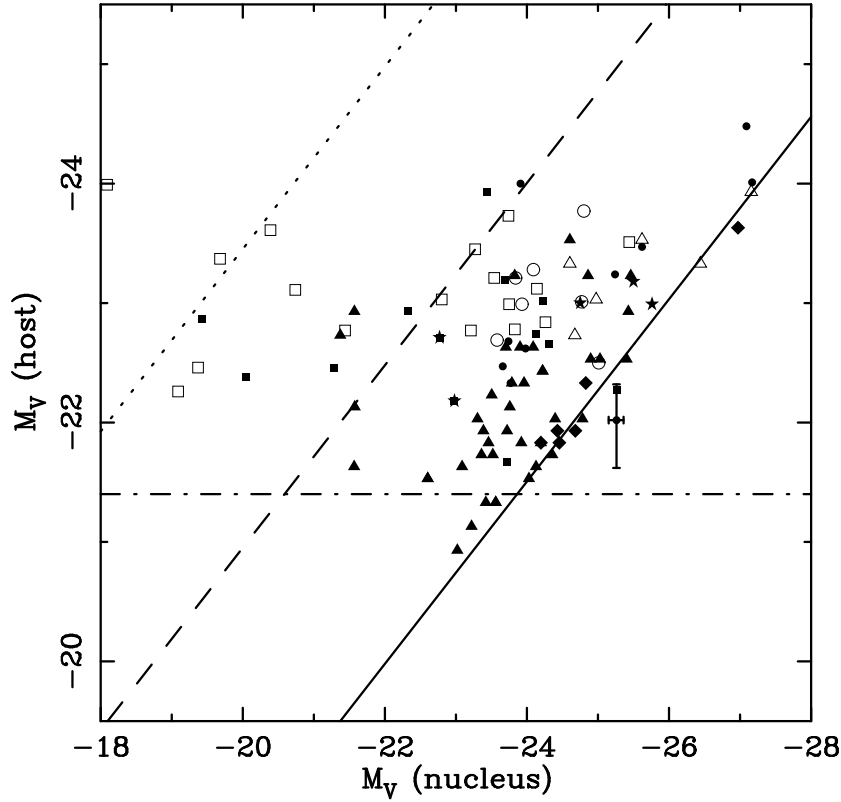}
\caption{Host absolute magnitude plotted versus nuclear absolute magnitude for
the quasars studied by Floyd et al. (2003; circles), Dunlop et al. (2003; 
squares), McLeod \& Rieke (1994; triangles), McLeod \& McLeod (2001; 
diamonds), and also for five objects reimaged with {\it HST}\ by Percival et 
al. (2001; stars).  The solid line illustrates the predicted limiting
relation on the assumption of Eddington-limited accretion, with the dashed
and dotted lines denoting 10\% and 1\% of the Eddington limit, respectively.
The one object in this combined sample that appears to be
more luminous than the Eddington limit is the luminous quasar 1252$+$020.
However, as indicated by the large error bars, this is also the object 
for which Floyd et al. (2003) have least confidence in the robustness
with which host and nuclear luminosity have been separated
(see Floyd et al. 2003 for further details).} 
\end{figure}

\section{Cosmological Evolution of the Black Hole-Spheroid Mass Ratio}

The two main results presented in the last two sections can be summarized 
as follows.
First, it is clear that the host galaxies of low-redshift quasars are normal
massive ellipticals. Second, it appears that by combining deep host galaxy
imaging with the spectroscopic H$\beta$ virial black hole mass estimator, 
low-redshift quasars can be used to provide an unbiased estimate of the 
black hole-spheroid mass ratio in the present-day inactive elliptical 
galaxy population.

These two results provide confidence that, through the study of quasars 
at higher redshifts, we can establish the cosmological evolution of the 
black hole-spheroid mass ratio in the general elliptical galaxy population.
This is important for two reasons. First, from a purely practical 
point of view, we are forced to study quasars to explore the redshift 
evolution of this mass relationship. This is simply because a virial mass 
estimator based on bright, observable, emission lines offers the only realistic
method by which to measure black hole masses in high-redshift objects.
Second, it can certainly be argued that, to all intents and purposes, the 
high-redshift elliptical galaxy population {\em is} the high-redshift quasar
population. Whereas only 1 in $10^4-10^5$ present-day ellipticals is 
active, Figure 1.6 coupled with a comparison of the present-day elliptical
and high-redshift quasar luminosity functions leads to the conclusion that 
at least 10\% of the progenitors of present-day massive ellipticals were
active quasars at $z \simeq 2.5$. 
 
So, to explore the cosmological evolution of the black hole-spheroid
mass ratio in massive galaxies, we require a version of the H$\beta$ 
virial mass estimator that can be applied to high-redshift quasars,
coupled with a means to estimate the masses of high-redshift quasar hosts.
Below I consider the current status of these two observational challenges, 
starting with the problem of black hole mass estimation at high redshift.

\subsection{Black Hole Mass Measurement in High-redshift Quasars}

The well-studied H$\beta$ emission line is observable from the ground out
to $z \simeq 3$. However, because it is redshifted 
into the near-infrared at a redshift of $z\sim1$, it is observationally
expensive to use H$\beta$ to estimate the black hole masses of $z>1$
quasars. Consequently, a concerted effort has recently been invested to 
establish whether or not any of the ultraviolet (UV) emission lines, so 
prominent in the observed optical spectra of high-redshift quasars, can be 
exploited and trusted to yield a comparably accurate and unbiased estimate
of black hole mass.

Two studies have recently been published that provide evidence that this can 
indeed be achieved. First, Vestergaard (2002) has proposed and calibrated 
a UV black hole mass estimator based on the FWHM of the C~IV 
emission line ($\lambda = 1549$\AA) and the continuum luminosity at 1350\AA. Second, 
McLure \& Jarvis (2002) have proposed and confirmed the robustness of a
UV black hole mass estimator based on the FWHM of the Mg~II 
emission line ($\lambda = 2799$\AA) and the continuum luminosity at 
3000\AA.

In terms of accessible redshift range, these two proposed mass estimators
are reasonably complementary, and in the near future it will be interesting 
to see how well they can be bootstrapped together to explore the 
black hole-spheroid mass ratio over a broad baseline in redshift.
However, at present it is probably fair to say that while the C~IV-based
estimator in principle allows black hole mass estimation from optical
spectroscopy out to $z \simeq 5$, the Mg~II-based estimator appears to be 
more robust and is better understood. 

The main reason for adopting Mg~II as the UV tracer of BLR velocity is
that, like H$\beta$, Mg~II is a low-ionization line. Furthermore,
due to the similarity of their ionization potentials, it is reasonable 
to expect that the Mg~II and H$\beta$ emission lines are 
produced by gas at virtually the same radius from the central
ionizing source. Although care has to be taken in dealing with Fe~II 
contamination in the vicinity of the Mg~II line, this 
presumption has now been directly tested and confirmed 
by McLure \& Jarvis (2002) through a comparison of Mg~II FWHM and 
H$\beta$ FWHM for a sample of 22 objects with
reverberation mapping results for which it also proved possible to
obtain Mg~II FWHM measurements. 

Building on this result, McLure \& Jarvis (2002) have produced a calibrated,
reliable, Mg~II virial black hole mass estimator that can be applied 
over the redshift range $0.3<z<2.5$ from straightforward optical spectroscopy.
In terms of a useful formula the final calibration of this UV 
black hole mass estimator is given by McLure \& Jarvis as:
 
\begin{displaymath}
\frac{ M_\bullet} {\mbox{\,$M_{\odot}$}}  =3.37\left(\frac{\lambda
L_{3000}}{10^{37}\,{\rm W}}\right)^{0.47}\left(\frac{{\rm FWHM(Mg~II})}
{{\rm km\,s}^{-1}}\right)^{2}
\end{displaymath}

\begin{figure}
\psfig{file=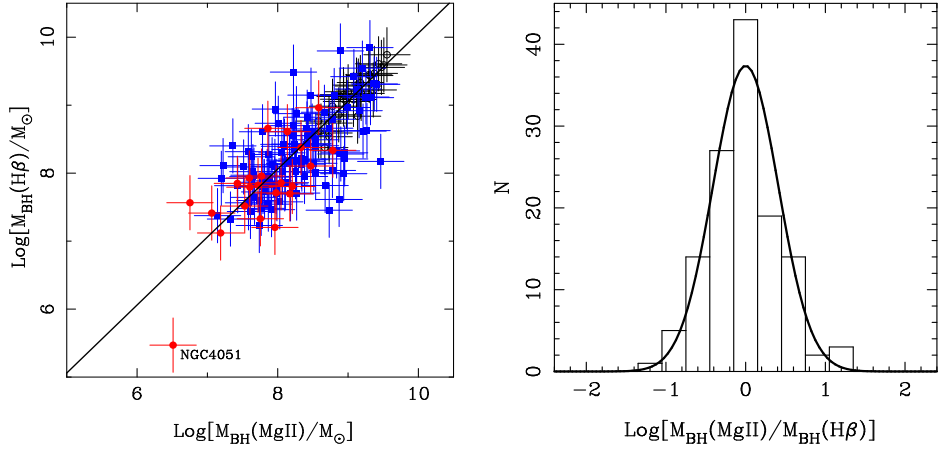,width=12.5cm,angle=0}     
\caption{{\it Left:}\ The optical (H$\beta$) versus UV (Mg~II) virial 
black hole estimators for 150 objects from the combined RM (filled circles), 
LBQS (filled squares), and MQS (open circles) samples described by McLure \& 
Jarvis (2002). The solid line is the BCES bisector fit to the 128 objects 
from the MQS and LBQS samples and has a slope of $1.00\pm0.08$. The outlying 
narrow-lined Seyfert NGC 4051 has been highlighted. {\it Right:}\ Histogram 
of $\log M_\bullet({\rm Mg~II})-\log M_\bullet({\rm H}\beta)$ for 
the 128 objects from the LBQS and MQS samples. Also shown is the 
best-fitting Gaussian, which has $\sigma=0.41$.}
\end{figure}

The robustness of this new black hole mass estimator is illustrated in 
Figure 1.7.
The left-hand panel 
shows a comparison of the results derived from the established 
optical (H$\beta$) black hole mass estimator plotted
against the results from this  new UV (Mg~II) 
black hole mass estimator for a 
combined sample of
150 objects [see McLure \& Jarvis (2002) for sample details].
Also shown is the BCES 
bisector fit to the data, which has the form 
\begin{displaymath}
\log M_\bullet({\rm H}\beta)=1.00(\pm0.08)\log M_\bullet({\rm Mg~II})+0.06(\pm0.67),
\end{displaymath}
\noindent
perfectly consistent with a linear relation. The right-hand panel shows a 
histogram of 
$\log M_\bullet({\rm Mg~II})-\log M_\bullet({\rm H}\beta)$ for this quasar 
sample.  The solid line shows the
best-fitting Gaussian, which has $\sigma=0.41$. These results 
lead McLure \& Jarvis to conclude that, 
compared to the traditional optical black hole mass estimator, the 
new UV estimator provides results which are unbiased and of equal
accuracy. 

In concluding this subsection I note that an exciting demonstration of how, 
through near-infrared spectroscopy,  this estimator can be applied 
to estimate the black hole masses of the most distant known quasars at $z > 6$
has recently been provided by Willott, McLure, \& Jarvis (2003). 

\subsection{Host Galaxy Mass Measurement at High Redshift}

The price to be paid for having a bright quasar nucleus from which to 
make emission-line based black hole mass estimates at high redshift 
is, of course, that the measurement of the mass of the host galaxy becomes
a challenge. The combination of generally unfavorable $K$ corrections
and strong surface brightness dimming means that the effective study
of quasar hosts beyond $z \simeq 1$ is much harder than the study of 
low-redshift hosts discussed above.

It is thus natural and sensible to consider whether there is any 
alternative to host galaxy imaging that might be utilized to estimate 
host galaxy mass at high redshift. This is the motivation behind the recent
suggestion by Shields et al. (2003) that the FWHM of the [O~III] emission
line in high-redshift quasars provides a measure of the velocity dispersion
of the stars in the central regions of the host galaxy. If true, then 
the black hole-spheroid mass ratio in high-redshift quasars could be estimated
simply from high-quality optical and near-infrared spectroscopy.

Unfortunately, this claim seems to be optimistic. While there is reasonable
evidence that [O~III] FWHM can be used as a proxy for central stellar
velocity dispersion in low-luminosity, low-redshift AGNs (Nelson 2000), 
there are 
many reasons why this is unlikely to be the case in more luminous objects,
especially at high-redshift (Boroson 2003). 
Moreover, as demonstrated in Figure 1.8, 
there is little evidence of a statistically useful correlation between 
the proposed [O~III] estimator of spheroid mass and spheroid luminosity
for the quasars and AGNs considered earlier in Figure 1.5.
Unfortunately, therefore, to obtain meaningful estimates of host galaxy 
masses for high-redshift quasars there currently appears to be no 
alternative but to attempt to measure host galaxy luminosities.

\begin{figure}
\psfig{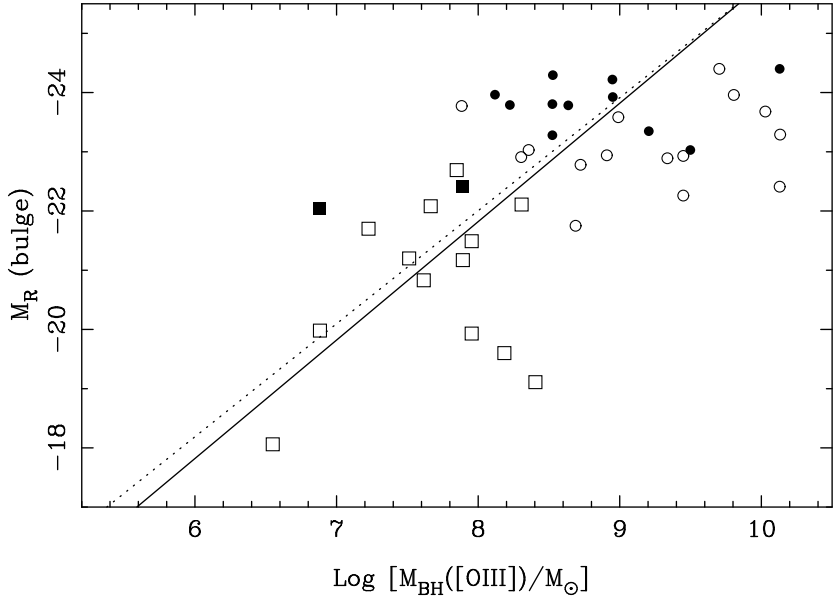}     
\caption{Host spheroid absolute magnitude plotted against the [O~III]-based 
mass estimator proposed by Shields et al. (2003) for the subset of 
quasars and Seyferts shown in Figure 1.5 for which a reliable FWHM for
[O~III] could be measured. Comparison of this ``relationship'' with the
tight correlation shown in Figure 1.5 provides little confidence that
[O~III] can be used as a reliable estimator of stellar velocity dispersion
in the hosts of high-redshift quasars. Removal of the radio-loud objects
(black data points) does not improve the significance of the 
correlation.}
\end{figure}

What, then, are the prospects for determining the masses of high-redshift 
quasar host galaxies from deep imaging data? Obviously high angular
resolution and a sound knowledge of the detailed form of the point-spread 
function remain a necessity. Also, to minimize the uncertainty 
in galaxy mass estimation introduced by the evolution of the host stellar
population, it is desirable to undertake observations of high-redshift
quasar hosts at near-infrared wavelengths.

The advent of the NICMOS camera on {\it HST}\ allowed its
unique ability to provide images with robust and repeatable point-spread 
functions to be extended into the near-infrared. Although NICMOS is only
effective out to the $H$ band at 1.6 $\mu m$, this was sufficient to allow 
Kukula et al. (2001) to extend the restframe $V$-band study of the hosts of 
moderate luminosity quasars from the $z = 0.2$ regime probed by Dunlop 
et al. (2003) out to $z \simeq 2$.

Kukula et al. (2001) defined two new quasar samples at $z\simeq 1$ and $2$,
confined to the luminosity range $-24 > M_V > -25$ mag, and by observing these
with NICMOS through the $J$ and $H$ band, respectively, 
obtained line-free images of the quasar hosts at both redshifts 
in the restframe $V$ band.

At $z \simeq 1$ Kukula et al. found it was still possible, on the basis of 
the NICMOS data, to estimate the scale lengths of the host galaxies with 
sufficient accuracy to demonstrate that they were at least consistent 
with the (passively evolved) Kormendy relation derived by Dunlop et al. at 
$z \simeq 0.2$ (see Fig. 1.4).
Therefore, just as at low redshift, the host galaxies of quasars at
$z\simeq1$ appear to be large, luminous systems. However, while for 
three of the $z \simeq 1$ quasars Kukula et al. found 
strong evidence that the hosts follow a de Vaucouleurs surface
brightness profile, in the majority of cases
the data did not allow an unambiguous fit. 

By $z \simeq 2$, despite deliberately deeper imaging, the increased
size of the $H$-band point-spread function, coupled presumably with the
impact of additional surface brightness dimming, meant that Kukula et al.
were unable to determine unambiguous morphologies for any host galaxy,
and only highly uncertain scale lengths in most cases. However, extended 
starlight was still detected in every object and it proved possible
to still obtain meaningful measurements of host luminosity.

The most robust result from this study that can be extracted at all redshifts
is the average luminosity of the quasar host galaxies in the restframe
$V$ band. This is plotted against redshift in Figure 1.9, which shows that,
under the assumption of passive evolution, the hosts of comparably luminous
quasars are basically unchanged in mass out to $z \simeq 2$.

\begin{figure}
\psfig{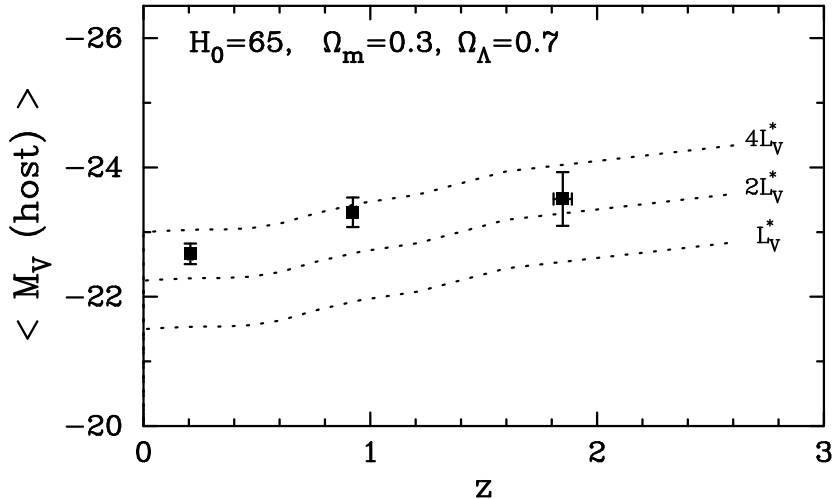}     
\caption{Mean absolute magnitude versus redshift for the quasar host galaxies 
from the HST imaging programs of McLure et al. (1999), 
Kukula et al. (2001) and Dunlop et al. (2003). The dotted lines show the 
passive evolutionary tracks for present-day $L^{\star}$, $2\,L^{\star}$, and
$4\,L^{\star}$ galaxies assuming that they formed in a single starburst event 
at $z\approx 5$. At all redshifts the average mass of the quasar host galaxies 
is consistent with that of a present-day $3\,L^{\star}$ elliptical, under
the assumption of passive evolution.}
\end{figure}

Although Figure 1.9 represents an interesting first attempt to determine
the mass evolution of quasar hosts, this result cannot yet be regarded as 
anything like as secure as the results on low-redshift quasar hosts presented
in \S~1.2. First, the samples are small. Second, at present the mass
estimates remain vulnerable to the validity of assuming passive evolution.
Third, while the complete quasar host sample appears, on average, to have the 
same stellar mass from $z \simeq 0.2$ to $z \simeq 2$, there is evidence 
within the data that the RQQ hosts are less massive at 
high redshift by a factor of $2-3$. This is consistent with, although
less extreme than, the decrease in radio-quiet host galaxy
mass with increasing redshift reported by Ridgway et al. (2001) from an 
analysis of NICMOS data. 
Recently, Lacy et al. (2002) have also found evidence for 
a slight decline in host galaxy mass by $z \simeq 1$, using ground-based
near-infrared imaging coupled with active/adaptive optics.

\section{Future Prospects}

To obtain improved estimates of the masses of high-redshift 
quasar hosts will require color information (to test the assumption
of passive evolution; Kukula et al. 2003) and the extension of 
high-resolution imaging observations into the $K$-band with the largest 
available telescopes. This work is already underway  
(e.g., Hutchings 2003) and over the next few years,
with careful study design (e.g., selection of quasars within a few arcseconds
of an appropriate star for reliable PSF determination) and the necessary 
major investment of telescope time, it is not unreasonable to expect 
that the Gemini telescopes and the VLT can 
revolutionize the effective study of 
high-redshift quasar hosts in much the same way as HST has revolutionized the
study of low-redshift hosts.

In the very near future the new UV viral black hole 
mass estimators described above will be applied to the extensive databases
of quasar optical spectra now being released by, for example, the Sloan Digital
Sky Survey.

Therefore, within the next 2 years or so, it is not unreasonable to anticipate
the construction of the first robust measurements of the redshift dependence 
of black hole-spheroid mass ratio within the bright quasar population out 
to $z \simeq 5$. Such measurements promise to provide fundamental new 
insights into our understanding of the relationship between black hole 
and galaxy formation.

\end{document}